\documentclass[
    ,final            
    ,sort&compress]
  {aipproc}
\layoutstyle{8x11single}

\usepackage{amsmath}

\begin{document}

\title{Exact Methods for Self Interacting Neutrinos}


\author{Y. Pehlivan}{
  address={Mimar Sinan GS{\"{U}}, Department of Physics, {\c{S}}i{\c{s}}li, {\.{I}}stanbul 34380, Turkey}
}

\author{A.~B. Balantekin}{
  address={Department of Physics, University of Wisconsin - Madison, Wisconsin 53706 USA} 
}

\author{Toshitaka Kajino}{
  address={National Astronomical Observatory of Japan 2-21-1 Osawa, Mitaka, Tokyo, 181-8588, Japan}
 ,altaddress={Department of Astronomy, University of Tokyo, Tokyo 113-0033, Japan}
}

\begin{abstract} 
The effective many-body Hamiltonian which describes vacuum oscillations
and self interactions of neutrinos in a two flavor mixing scheme under the single
angle approximation has the same dynamical symmetries as the well known BCS pairing
Hamiltonian. These dynamical symmetries manifest themselves in terms of a set of constants
of motion and can be useful in formulating the collective oscillation modes in an
intuitive way. In particular, we show that a neutrino spectral split can be simply viewed
as an avoided level crossing between the eigenstates of a mean field Hamiltonian which
includes a Lagrange multiplier in order to fix the value of an exact many-body constant of
motion. We show that the same dynamical symmetries also exist in the three neutrino
mixing scheme by explicitly writing down the corresponding constants of motion. 
\end{abstract}

\maketitle

Neutrinos are produced in copious amounts by various astrophysical sources. For example, a
core collapse supernova releases $\% 99$ of the gravitational binding energy of the
pre-supernova in the form of neutrinos \citep[see Refs.][for
review]{Kotake:2005zn,Woosley:2002zz,barkat,Wheeler:2002gw}. These neutrinos are 
excellent probes into the physics of the supernova and believed to play a role in the supernova
dynamics as well as the subsequent r-process nucleosynthesis. Black hole accretion disks are
also likely to be major sources of neutrinos \citep{Matteo:2002ck,Caballero:2009ww}. In
the Early Universe neutrinos were produced abundantly and influenced the Big Bang
nucleosynthesis \citep[see Refs.][for review]{Dolgov:2002wy,Hannestad:2006zg}. 

Determining the impact neutrinos requires a careful study of their energy distribution.
Although the initial energy distribution can be given with a specific model, it is
subsequently modified as the neutrinos undergo flavor evolution subject to the refractive
effects of the background matter. The refraction of neutrinos due to the other particles
in the background such as protons, neutrons and electrons, is proportional to the forward
scattering amplitude since it is only in the forward direction that the scattering
amplitudes add up coherently. This gives rise to the well known MSW effect in the Sun.
But if the neutrino density is sufficiently high, then the self refraction of neutrinos
can also make a significant contribution to the flavor evolution as is the case for the
core collapse supernovae \citep{savage&malaney}, the Early Universe \citep{fuller&mayle}
and in black hole accretion disks \citep{Malkus:2012ts}. When neutrinos scatter off each
other, those diagrams in which neutrinos completely exchange their momenta also add up
coherently in addition to the forward scattering diagrams \citep{Pantaleone:1992eq}. The
exchange diagrams couple neutrinos with different energies and turn flavor evolution of
the system into a nonlinear many-body phenomenon. A rich set of flavor oscillation modes
arises due to this nonlinearity which is a subject of intense study in the recent years.
In particular, collective flavor oscillation modes in which neutrinos of different
energies oscillate with the same frequency were identified and thoroughly studied in terms
of their astrophysical implications
\citep{Fuller:2005ae,Kostelecky:1994dt,Samuel:1995ri,Friedland:2003dv,Friedland:2003eh,
Duan:2005cp,Friedland:2006ke,Duan:2006an,Hannestad:2006nj,Raffelt:2007cb}. 

Salient features of collective flavor oscillations can be captured by a simplified model
in which neutrinos undergo vacuum oscillations and self interactions in the absence of a
net leptonic background under the so-called single angle approximation. It was pointed
out in Ref. \citep{Pehlivan:2011hp} that, in the case of a two neutrino mixing scheme,
this model has the same dynamical symmetries as the reduced pairing Hamiltonian which is
used in the context of the BCS model of superconductivity to describe the electron pairs
in the conduction band of a metal and also in the context of the nuclear shell model to
describe the nucleon pairs occupying the valance shell of a nucleus. These dynamical
symmetries guarantee that the model is exactly solvable in both its original many-body
form and in the framework of the commonly used mean field approximation.

The goal of this contribution is to emphasize that a thorough examination of the symmetries and
exact solutions of the self interacting neutrinos would be helpful in developing a deeper
insight into the nature of the collective flavor oscillations. Such a study is naturally
complementary to the numerical techniques that have been developed and successfully
applied to the problem so far. As an example, we will consider the adiabatic flavor
evolution of neutrinos as they radiate from a source and undergo spectral splits by
exchanging parts of their spectra. This behavior was first observed in the numerical
simulations of the system under the mean field approximation \cite[see Ref.][for a
review]{Duan:2010bg}. They were analytically explained in terms of the adiabatic time
evolution of the instantaneously stable mean field configurations viewed from a rotating
frame of reference in the neutrino isospin space \citep{Raffelt:2007cb}. Here we offer an
alternative view of a spectral split as an avoided level crossing between the eigenstates
of a mean field Hamiltonian which includes a Lagrange multiplier in order to fix the value
of an exact many-body constant of motion which cannot be otherwise fixed in the mean field
approximation scheme \citep{Pehlivan:2011hp}. 

Note that the dynamical symmetries and the corresponding constants of motion of the self
interacting neutrinos were so  far examined only for two flavor mixing. It is natural to
ask if similar symmetries also exist for three flavor mixing. We show
that the answer is positive by explicitly writing down the constants of
motion of the exact many-body Hamiltonian describing the vacuum oscillations and self
interactions of three neutrino flavors in the single angle approximation. The
implications of these symmetries, including their roles in the multiple spectral splits 
are subject to further study and will be reported elsewhere. 

We start by formulating the problem for two mixing flavors. We take them to be $\nu_e$ and
an orthogonal flavor that we denote by $\nu_x$ which can be either $\nu_{\mu}$ or
$\nu_{\tau}$ or a normalized combination of them. The particle operator for a neutrino of
flavor $\alpha=e,x$ with momentum $\mathbf{p}$ is denoted by
$a_{\alpha}\left(\mathbf{p}\right)$. Typically, additional quantum numbers besides the
momentum are needed to distinguish the neutrinos but we choose to keep our formulas simple
by not explicitly displaying them in our notation. Instead, $\mathbf{p}$ can be viewed as
a multiple index like $(\mathbf{p},s_1,s_2,\dots)$. 

It is useful to introduce the \emph{isospin operator} $\vec{J_\mathbf{p}}=(J_\mathbf{p}^+, J_\mathbf{p}^-,
J_\mathbf{p}^3)$ whose components are given by 
\begin{equation} 
J_{\mathbf{p}}^+= a_e^{\dagger}(\mathbf{p})a_x(\mathbf{p})~,\qquad 
J_{\mathbf{p}}^-= a_x^{\dagger}(\mathbf{p})a_e(\mathbf{p})~,\qquad 
J_{\mathbf{p}}^3=\frac{1}{2}\left(a_e^{\dagger}(\mathbf{p})a_e(\mathbf{p})-a_x^{\dagger}(\mathbf{p})a_x(\mathbf{p})\right)~. 
\label{Flavor Isospin Operators} 
\end{equation} 
Note that we use boldface letters to indicate vectors in momentum space (e.g.
$\mathbf{p}$) and arrows to indicate vectors in flavor space (e.g. $\vec{J}$). The
components of the isospin operator obey the $SU(2)$ commutation relations    
\begin{equation} 
[J_{\mathbf{p}}^+,J_{\mathbf{q}}^-]=2 \delta_{\mathbf{p}\mathbf{q}}J_{\mathbf{p}}^3~,\qquad 
[J_{\mathbf{p}}^3,J_{\mathbf{q}}^{\pm}]=\pm \delta_{\mathbf{p}\mathbf{q}}J_{\mathbf{p}}^{\pm} 
\label{Flavor Isospin Algebra} 
\end{equation} 
such that one has as many mutually commuting $SU(2)$ \emph{isospin algebras} as the number of
neutrinos. It follows from Eq. (\ref{Flavor Isospin Operators}) that each isospin algebra
is realized in the spin-$1/2$ representation and that the electron neutrino is 
isospin up. 

The particle operators in the mass basis are denoted by $a_i\left(\mathbf{p}\right)$ where
$i=1,2$ indicates the eigenstate with mass $m_i$. The transformation from the flavor basis into mass
basis is a global rotation in the sense that it is the same for neutrinos of all energies.
One can also express the isospin operator in terms of its components in the mass basis,
i.e., $\vec{J_\mathbf{p}}=(\mathcal{J}_\mathbf{p}^+, \mathcal{J}_\mathbf{p}^-,
\mathcal{J}_\mathbf{p}^3)$ where
\begin{equation} 
\mathcal{J}_{\mathbf{p}}^+= a_1^{\dagger}(\mathbf{p})a_2(\mathbf{p})~,\qquad 
\mathcal{J}_{\mathbf{p}}^-= a_2^{\dagger}(\mathbf{p})a_1(\mathbf{p})~,\qquad 
\mathcal{J}_{\mathbf{p}}^3=\frac{1}{2}\left(a_1^{\dagger}(\mathbf{p})a_1(\mathbf{p})-a_2^{\dagger}(\mathbf{p})a_2(\mathbf{p})\right)~. 
\label{Mass Isospin Operators} 
\end{equation} 
In order to avoid confusion, we use curly letters $\mathcal{J}^a$ to denote the components
of isospin operator in the mass basis. They satisfy the same commutation relations as
those given in Eq. (\ref{Flavor Isospin Algebra}). 

At this point we introduce our summation convention for all isospin operators 
as follows:
\begin{equation} 
\vec{J}_p\equiv\sum_{|\mathbf{p}|=p}\vec{J}_{\mathbf{p}} \qquad \mbox{and} \qquad 
\vec{J}\equiv\sum_p \vec{J}_p~. 
\label{Totals} 
\end{equation}
Here, $\vec{J}_p$ represents the total isospin operator of all neutrinos with the same
energy $p$ and $\vec{J}$ represents the total isospin operator of all neutrinos.
Since the operators $\vec{J}_p$ and $\vec{J}$ are sums of individual $SU(2)$ operators,
their components also obey the $SU(2)$ commutation relations. 

The Hamiltonian describing the vacuum oscillations of neutrinos can be written as 
\begin{equation} 
H_{\nu}=\sum_{\mathbf{p}}\left(\frac{m_1^2}{2p}a_1^{\dagger}(\mathbf{p})a_1(\mathbf{p})
+\frac{m_2^2}{2p}a_2^{\dagger}(\mathbf{p})a_2(\mathbf{p})\right) 
=\sum_{p}\omega_p\vec{B}\cdot\vec{J}_p~. 
\label{Vacuum Oscillation Hamiltonian} 
\end{equation} 
Here $\omega_p=({m_2^2-m_1^2})/{2p}$ is the vacuum oscillation frequency of the neutrino
with energy $p$. Note that the summation convention introduced in Eq. (\ref{Totals}) is used for
neutrinos with the same energy. In Eq. (\ref{Vacuum Oscillation Hamiltonian}), $\vec{B}$
is a vector which points in the negative direction along the third axis in the mass basis.
Its components are given by 
\begin{equation} 
\vec{B}=(0,0,-1)_{\mbox{\tiny mass}}=(\sin2\theta,0,-\cos2\theta)_{\mbox{\tiny flavor}}~ 
\end{equation} 
in mass and flavor bases, respectively. We also note that the equality in Eq. (\ref{Vacuum
Oscillation Hamiltonian}) is correct up to a term proportional to identity 
which can always be subtracted from the Hamiltonian.

The effect of scattering on the neutrino flavor oscillations in matter can be
described by an effective Hamiltonian which takes into account only those terms which add
up coherently over the scatterers. In the case of the neutrino-neutrino scattering, only  
forward scattering diagrams in which there is no momentum transfer between the particles, and 
exchange diagrams in which particles swap their momenta add up
coherently. As a result, the effective Hamiltonian which describes self interactions of
neutrinos have the following form \citep{Sigl:1992fn,Sawyer:2005jk}:
\begin{eqnarray} 
H_{\nu\nu}=\frac{\mu}{2}\sum_{\mathbf{p},\mathbf{q}} 
\left[\right.a_e^{\dagger}(\mathbf{p})a_e(\mathbf{p})a_e^{\dagger}(\mathbf{q})a_e(\mathbf{q})+a_x^{\dagger}(\mathbf{p})a_x(\mathbf{p})a_x^{\dagger}(\mathbf{q})a_x(\mathbf{q})
+a_x^{\dagger}(\mathbf{p})a_e(\mathbf{p})a_e^{\dagger}(\mathbf{q})a_x(\mathbf{q}) 
+a_e^{\dagger}(\mathbf{p})a_x(\mathbf{p})a_x^{\dagger}(\mathbf{q})a_e(\mathbf{q}))\left.\right]~.
\label{Self Interaction Hamiltonian} 
\end{eqnarray} 
Here $\mu=\sqrt{2}G_F/V$ where $V$ denotes the quantization volume. 
Note that the Hamiltonian in Eq. (\ref{Self Interaction Hamiltonian}) is valid within the
so called single angle approximation in which the dependence of the scattering amplitudes
on the angle between the propagation direction of neutrinos is ignored. It can be shown
that the Hamiltonian in Eq. (\ref{Self Interaction Hamiltonian}) is equal to 
\begin{equation}
\label{self interactions}
H_{\nu\nu}=\mu\vec{J}\cdot\vec{J}
\end{equation}
up to some terms which are proportional to identity. The Hamiltonian in Eq. (\ref{self
interactions}) has the same form in both the mass and flavor bases 
because these bases are related by a global rotation which leaves all scalar products
invariant.

The total Hamiltonian describing the self interactions of neutrinos together with 
vacuum oscillations is given by the sum of the terms in Eqs. (\ref{Vacuum Oscillation Hamiltonian})   
and (\ref{self interactions}), i.e., 
\begin{equation} 
H=\sum_p\omega_p\vec{B}\cdot\vec{J}_p+\mu\vec{J}\cdot\vec{J}~. 
\label{total hamiltonian} 
\end{equation}
This Hamiltonian belongs to one of the three classes of exactly solvable Hamiltonians
which were first systematically studied by Gaudin in 1976 \citep{Gaudin:1976,Gaudin:1983}.
Gaudin was interested in finding integrable Hamiltonians which describe spin systems in an
external magnetic field with long range spin-spin interactions. He classified
those integrable Hamiltonians that he identified into three classes which are known as the rational,
trigonometric and elliptic models. The Hamiltonian in Eq. (\ref{total hamiltonian})
belongs to the class of rational  models. As will be discussed in more detail below, it
has as many constants of motion (or invariants) as the number of energy modes in the system. These
constants of motion were identified by Gaudin and are known as the ``rational Gaudin
magnet Hamiltonians.'' It should be noted that although Gaudin was interested in real
spins as opposed to isospins that appear in Eq. (\ref{total hamiltonian}), this
distinction is not important as long as the integrability and the exact solutions are
concerned because both the real spins and the isospins obey the same algebra given in Eq.
(\ref{Flavor Isospin Algebra}). 

In the language of neutrino isospin, the invariants identified by Gaudin are
given by
\begin{equation} 
\label{Invariants} 
h_p=\vec{B}\cdot\vec{J}_p+2\mu\sum_{q\left(\neq
p\right)}\frac{\vec{J}_p\cdot\vec{J}_q}{\omega_p-\omega_q}~. 
\end{equation} 
It is straightforward to show that these operators commute with one another 
and with the neutrino Hamiltonian: 
\begin{equation} 
\left[{h}_p,{h}_q\right]=0\qquad\mbox{and}\qquad\left[{H},{h}_p\right]=0 
\qquad\mbox{for all $p$ and $q$.}\qquad
\label{Conservation of GMH} 
\end{equation} 
It is possible to express the invariants of the model in a number of alternative ways
using linear or nonlinear combinations of those given in Eq. (\ref{Invariants}). In
particular, their sum is simply equal to 
\begin{equation}
\label{sum}
\sum_p h_p=-\mathcal{J}^3=\frac{N_2-N_1}{2}
\end{equation}
where $N_i$ is the total number of neutrinos in the $i^{\mbox{\footnotesize th}}$ mass
eigenstate. Note that the total number of neutrinos, $N_1+N_2$, is also constant because
we consider only the vacuum oscillations and scatterings of the neutrinos. Therefore Eq.
(\ref{sum}) tells us that $N_1$ and $N_2$ are individually conserved. 

Another case where the constants of motion take a simple form is the $\mu\to 0$ limit
where all neutrino-neutrino interactions cease. In this limit, we have 
\begin{equation}
\label{Invariants limit} 
\lim_{\mu\to 0}h_p=-\mathcal{J}_p^3=\frac{n_2(p)-n_1(p)}{2}
\end{equation} 
where $n_i(p)$ is the total number of neutrinos in the $i^{\mbox{\footnotesize th}}$ mass
eigenstate with energy $p$. Note that the total number of neutrinos in a given energy
mode, $n_1(p)+n_2(p)$, is conserved for any value of $\mu$ because neutrinos either keep
their momenta or exchange it in the current model.  Therefore, Eq. (\ref{Invariants
limit}) simply expresses the fact that $n_1(p)$ and $n_2(p)$ are individually conserved in
$\mu\to 0$ limit. However, away from the $\mu\to 0$ limit and except for the combination
in Eq.  (\ref{sum}), the constants of motion in Eq. (\ref{Invariants}) are nontrivial and
cannot be expressed in terms of the neutrino number operators.
  
It should also be noted that the operators in Eq. (\ref{Invariants}) are invariant only
under the \emph{ideal} conditions, i.e., when  the single angle approximation is adopted,
there is no \emph{net} leptonic background, and the volume occupied by the neutrinos is
fixed ($\mu=$ constant).  However, the constants of motion may still be useful away from
these ideal conditions. For example, one can decompose the Hamiltonian into ideal and
non-ideal parts as 
\begin{equation}
H=H_{\mbox{\footnotesize ideal}}+H_{\mbox{\footnotesize non-ideal}}~.
\end{equation}
In this case, the time evolution of the ``constants of motion'' will only be due to the
non-ideal part, i.e.,  
\begin{equation}
\frac{d}{dt}h_p=-i[h_p,H_{\mbox{\footnotesize non-ideal}}]~.
\end{equation}
because they commute with the ideal part. Therefore the invariants can provide a
convenient set of variables subject to a simpler time evolution. 

It is worth mentioning that the Hamiltonian in Eq. (\ref{total hamiltonian}) was studied
and its integrability was already known before Gaudin's work. In fact it was first
introduced in 1957 by Bardeen, Cooper and Schrieffer in order to describe the pairing of
valance electrons in a superconductor \citep{Bardeen:1957mv}. In the context of electron
pairs, the role of the Gaudin's spins or the neutrino isospin is played by \emph{pair
quasi-spin} operator. The (reduced) BCS pairing Hamiltonian is given by
\begin{equation} 
H_{\mbox{\tiny BCS}}=\sum_k2 \epsilon_k{t}_k^3-G{T}^+{T}^-~. 
\label{BCS Hamiltonian} 
\end{equation} 
It describes a set of spin up ($c_{k\uparrow}$) and spin-down ($c_{k\downarrow}$)
electrons (Cooper pairs) which can occupy a set of single particle energy levels denoted
by $\epsilon_k$. The components of the quasi-spin operator $\vec{t}_k$ are given by    
\begin{equation} 
t_k^+=c_{k\uparrow}^\dagger c_{k\downarrow}^\dagger~, \qquad  
t_k^-=c_{k\downarrow} c_{k\uparrow}~ \qquad\mbox{and}\qquad 
t_k^3=\frac{1}{2}\left(c_{k\uparrow}^\dagger c_{k\uparrow}+c_{k\downarrow}^\dagger c_{k\downarrow}-1\right) 
\label{Quasi-spin Operators} 
\end{equation}
and they obey the same $SU(2)$ commutation relations as those given in Eq. (\ref{Flavor
Isospin Algebra}). In the quasi-spin scheme, a single particle level $\epsilon_k$ has
quasi-spin up if it is occupied by a pair and quasi-spin down if it is not.
$\vec{T}=\sum_k \vec{t}_k$ denotes the total quasi-spin of all levels and $G>0$ is the
pairing strength. The BCS pairing Hamiltonian is also used in nuclear shell model to
describe pairing between the nucleons in the valance shell. 

It is easy to see that in the mass basis where $\vec{B}=(0,0,-1)$, the neutrino
Hamiltonian in Eq. (\ref{total hamiltonian}) has the same form as the BCS pairing
Hamiltonian in Eq. (\ref{BCS Hamiltonian}) up to an overall minus sign and a term
proportional to $\mu \mathcal{J}^3(\mathcal{J}^3-1)$. The overall minus sign may have
dynamical consequences on the stability of some solutions but it is irrelevant for a
discussion of the symmetries and the resulting exact solvability of both models. The term
$\mu \mathcal{J}^3(\mathcal{J}^3-1)$ is also unimportant in this context because
$\mathcal{J}^3$ is itself a constant of motion as was shown in Eq. (\ref{sum}). 

The exact solvability of the pairing model was first shown by Richardson in 1963
\citep{Richardson:1963} who found its exact eigenstates and eigenvalues using the method
of Bethe ansatz \citep{Bethe:406121}. This method gives analytical expressions for the
eigenstates and yields corresponding eigenvalues in terms of the roots of some algebraic
equations which are known as the Bethe ansatz equations. Although these equations still
call for a numerical  approach in generic cases, the resulting problem is significantly
less challenging than a brute force diagonalization of the Hamiltonian. In fact many
numerical and analytical techniques were developed to solve them, especially in the limit
of a large number of particles \cite{Richardson:1966zza,richardson:1802}. In recent
years, the problem of numerically solving the equations of Bethe ansatz received renewed
interest \citep{Roman:2002dh}, particularly in connection with the quench dynamics of
superconductors away from the stability \citep{Faribault:2011rv}. Pairing models and the
solutions of the related Bethe ansatz equations also receive attention in recent years due
to their connections with the conformal field theories and the matrix models
\citep{Asorey:2001wu,Jurco:2003pv}. Reviews can be found in Refs.
\citep{Sierra:2001cx,Dukelsky:2004re}.

A mean field type approximation is usually employed in the case of both self interacting
neutrinos and the BCS model. The exact solvability extends to the mean field case as was
shown by Yuzbashyan \emph{et al} who derived formal solutions of the resulting mean field
equations in the context of the BCS model \citep{yuzbashyan}. Collective modes of
behavior for Cooper pairs were also analyzed in the same reference. Note that two of these
collective modes were already known in the context of neutrinos as synchronized and
bipolar oscillations. More recently, all of these modes were identified and classified for
neutrinos in an independent study \citep{Raffelt:2011yb}. 

In the mean field approximation, neutrino-neutrino interactions are represented by an
effective one-body scheme in which each neutrino interacts with an average potential
created by all other neutrinos. One way to implement this approximation is to employ the
operator product linearization through which the quadratic operator $\vec{J}_p\cdot
\vec{J}_q$ is approximated as    
\begin{equation} 
\label{MF approximation} 
{\vec{J}_p}\cdot{\vec{J}_q} \sim  
{\vec{J}_p} \cdot \langle {\vec{J}_q} \rangle + \langle {\vec{J}_p} \rangle \cdot {\vec{J}_q} - 
\langle {\vec{J}_p} \rangle \cdot \langle {\vec{J}_q} 
\rangle~. 
\end{equation} 
Linearization of the neutrino evolution equations is also discussed in Ref. 
\citep{Vaananen:2013qja}.
The expectation values in the above equations should be calculated with respect to a state
which satisfies the condition $\langle {\vec{J}_p} \cdot {\vec{J}_q} \rangle = \langle
{\vec{J}_p} \rangle \cdot \langle {\vec{J}_q}\rangle$. This amounts the truncating the
Hilbert Space of the problem by excluding the entangled states because this condition is
satisfied only by the non-entangled states. These states are also the coherent states of
the orthogonal $SU(2)$ algebras presented in Eq.  (\ref{Flavor Isospin Operators})
\citep{Balantekin:2006tg}. The expectation values of the isospin operators, i.e., 
$\vec{P}_{\mathbf{p}} \equiv 2\langle\vec{J}_{\mathbf{p}}\rangle$, are called the
\emph{polarization vectors} where the  
factor of $2$ is included for convenience. The
polarization vectors are also subject to the summation rule introduced in Eq.
(\ref{Totals}). Application of the mean field approximation to the neutrino Hamiltonian
given in Eq. (\ref{total hamiltonian}) yields  
\begin{equation} 
\label{MF Hamiltonian} 
H\sim{H}^{\mbox{\tiny MF}}=\sum_p \omega_p\vec{B}\cdot\vec{J}_p+\mu\vec{P}\cdot\vec{J} 
\end{equation} 
where $\vec{P}$ is the total polarization vector which is the total potential that each
neutrino interacts with. This approximation is consistent only if the mean field evolves
in line with the evolution of the particles which collectively create it. In the
Heisenberg picture, this can be formulated by first calculating the quantum mechanical
equation of motion of the isospin operator from the mean field Hamiltonian, i.e.,
$d\vec{J}_p/dt=-i[\vec{J}_p, H^{\mbox{\tiny MF}}]$ and then taking the expectation values
of both sides \citep{Pehlivan:2011hp}. This yields the \emph{mean field consistency
equations}
\begin{equation} 
\label{MF EoM} 
\frac{d}{dt}\vec{P}_p=(\omega_p\vec{B}+\mu\vec{P})\times\vec{P}_p 
\end{equation} 
which should be satisfied for every momentum mode $p$. 

A straightforward calculation shows that the mean field Hamiltonian given in Eq. (\ref{MF
Hamiltonian}) does not commute with the exact many-body constants of motion given in Eq.
(\ref{Invariants}). However, their expectation values
\begin{equation} 
\label{Invariants MF} 
I_p \equiv 2\langle h_p \rangle=\vec{B}\cdot\vec{P}_p+\mu\sum_{q\left(\neq
p\right)}\frac{\vec{P}_p\cdot\vec{P}_q}{\omega_p-\omega_q} 
\end{equation} 
are still independent of time as can be directly verified from mean field consistency
equations. Non-conservation of the exact many-body constants of motion in the mean field
picture is particularly evident in the case of the invariant given in Eq. (\ref{sum})
because the occupation numbers of the first and second mass eigenstates clearly change as
the neutrinos interact with the mean field through the term
$\mathcal{P}^+\mathcal{J}^-+\mathcal{P}^-\mathcal{J}^+$ in Eq. (\ref{MF Hamiltonian}).
However, the average occupation numbers $\langle N_1 \rangle$ and $\langle N_2 \rangle$
continue to be invariant since $\mathcal{P}^3=2\langle\mathcal{J}^3\rangle$ is conserved.

The invariants of the neutrino Hamiltonian represent the dynamical symmetries of the
system which should be carefully studied in order to understand the collective modes of
behavior that the neutrinos display. Here, we would like to consider a particular example
concerning the adiabatic evolution of neutrinos from a region of high neutrino density
(like the surface of a proto-neutron star) into the vacuum. Numerical simulations of the
mean field equations under the relevant conditions showed that neutrinos completely
exchange parts of their spectra above or below a critical energy by the time they reach
the vacuum. This behavior is known as a spectral split and was analytically explained 
in terms of the instantaneously stable solutions of the  mean field consistency equations 
viewed from a rotating reference frame in the isospin space in such a way that
the rotation frequency of the frame yields the split frequency
\citep{Raffelt:2007cb,Raffelt:2007xt}. Here we consider this phenomenon from a different
perspective, namely as an avoided level crossing between the energy eigenvalues of the
mean field Hamiltonian which includes a Lagrange multiplier to fix the occupation numbers
$N_1$ and $N_2$ to chosen initial values. In this scheme, the value of the
Lagrange multiplier yields the split frequency. In what follows, we closely follow Ref.
\citep{Pehlivan:2011hp} where this approach was originally developed.

According to the adiabatic theorem, if a Hamiltonian varies slowly enough, then a system
which initially occupies one of its eigenstates evolves in such a way that it continues to
occupy the same \emph{instantaneous} eigenstate, as long as there is an energy gap between
this particular eigenstate and the others. One can easily find the eigenstates of the
mean field Hamiltonian given in Eq. (\ref{MF Hamiltonian}) but since the Hamiltonian
involves the mean field $\vec{P}$, its eigenstates necessarily involve $\vec{P}$ as a
parameter as well. On the other hand, these eigenstates should also satisfy the mean field
consistency condition $\vec{P}=2\langle \vec{J}\rangle$ and it is easy to see that this
condition restricts the value of the mean field $\vec{P}$. Since the component of
$\vec{P}$ along $\vec{B}$ is equal to $\langle N_1 \rangle-\langle N_2 \rangle$, this
tells us that, although the average total occupancies of mass eigenstates are conserved,
not all possible values are allowed for them in a steady state solution due to the self
consistency requirement of the mean field approximation. This is in contrast with the
exact many-body picture where the many-body Hamiltonian can be diagonalized
simultaneously with the number operators $N_1$ and $N_2$ and therefore there is a steady
state solution (i.e., an eigenstate of the many-body Hamiltonian) for all possible values
of these total occupancies. 

One way to accommodate any possible set of occupation numbers in the mean field picture
is to fix them by introducing a Lagrange multiplier $\omega_c$ before adopting the mean field
approximation as follows:
\begin{equation}
\label{lagrange}
\left(H+\omega_c\mathcal{J}^3_p\right)^{\mbox{\tiny MF}}
=-\sum_p(\omega_p-\omega_c)\mathcal{J}_p^3+\mu\vec{P}\cdot\vec{J}~. 
\end{equation}
This is equivalent to viewing the problem from a rotating reference frame in the isospin
space as pointed out in Refs. \citep{Raffelt:2007cb,Raffelt:2007xt}. In either case, the
extra degree of freedom $\omega_c$ can be used to set the desired occupation numbers. 

The instantaneous eigenstates of the Hamiltonian in Eq. (\ref{lagrange}) can be found with
the following transformation:
\begin{equation}
\label{alphas}
\begin{pmatrix}
\alpha_1(\mathbf{p}) \\ \alpha_2(\mathbf{p})
\end{pmatrix} =
\begin{pmatrix}
\cos{\theta_p}& e^{i\delta}\sin{\theta_p} \\ - e^{-i\delta}\sin{\theta_p}  & \cos{\theta_p} 
\end{pmatrix}
\begin{pmatrix}
a_1(\mathbf{p}) \\ a_2(\mathbf{p}) 
\end{pmatrix}~.
\end{equation}
Here $\theta_p$ and $\delta$ are given by 
\begin{equation}
\label{angles}
\sin{2\theta_p} =\sqrt{1-\frac{(\omega_c-\omega_p+\mu
\mathcal{P}^3)^2}{(\omega_c-\omega_p+\mu \mathcal{P}^3)^2+\mu^2
\mathcal{P}^+\mathcal{P}^-}}
\quad \mbox{and} \quad
e^{i\delta}=
\frac{\mathcal{P}^+}{|\mathcal{P}^+|}~.
\end{equation}
Note that here $\theta_p$ is not summed over all directions as defined in Eq.
(\ref{Totals}). It is indexed with $p$ because it does not depend on the direction of the
momentum. $\alpha_1(\mathbf{p})$ and $\alpha_2(\mathbf{p})$ are the instantaneous
\emph{non-interacting} degrees of freedom of the mean field Hamiltonian. In other words,
when the Hamiltonian in Eq. (\ref{lagrange}) is expressed in terms of them, it has the
form of a free Hamiltonian:  
\begin{equation}
\label{diagonalized}
\left(H+\omega_c\mathcal{J}^3\right)^{\mbox{\tiny MF}} =\sum_{\mathbf{p}} \lambda_p \left(
\alpha_1^\dagger(\mathbf{p})\alpha_1(\mathbf{p})-\alpha_2^\dagger(\mathbf{p})\alpha_2(\mathbf{p})\right)~.
\end{equation}
Here $\lambda_p$ is given by
\begin{equation} 
\label{lambda_w} 
\lambda_{p}=\frac{1}{2}\sqrt{(\omega_c-\omega_p+\mu\mathcal{P}^3)^2+\mu^2\mathcal{P}^+\mathcal{P}^-}~. 
\end{equation} 
At any given moment, instantaneous eigenstates of the Hamiltonian in Eq. (\ref{lagrange}) can be
written in terms of $\alpha_1(\mathbf{p})$ and $\alpha_2(\mathbf{p})$. For example, 
\begin{equation} 
\label{State} 
\prod_{\mathbf{p}}\alpha_1^{\dagger}(\mathbf{p})\;|0\rangle 
\end{equation} 
is a particular instantaneous eigenstate. The other eigenstates can be written similarly in
terms of $\alpha_1$'s and $\alpha_2$'s. 

It is easy to show that the new basis given in Eq. (\ref{alphas}) coincides with the
flavor basis when neutrinos occupy a very small volume ($V\to 0$, $\mu\to\infty$) and with
the mass basis when they occupy a very large volume ($V\to \infty$, $\mu\to 0$). If
initially there is more $\nu_e$ than $\nu_x$ in the system, then Eqs. (\ref{alphas}) and
(\ref{angles}) can be used to show that 
\begin{equation}
\label{high density limit}
\lim_{\mu\to\infty}\alpha_1(\mathbf{p})= a_e(\mathbf{p}) 
\qquad\mbox{and}\qquad
\lim_{\mu\to\infty}\alpha_2(\mathbf{p})= a_x(\mathbf{p})~. 
\end{equation}
If the opposite is true, i.e., initially there is more $\nu_x$ then $\nu_e$,
then $a_e$ and $a_x$ should be exchanged in the above equation.
In the limit where neutrino density approaches to zero, 
Eqs. (\ref{alphas}) and (\ref{angles}) give 
\begin{equation}
\label{vacuum limit}
\lim_{\mu\to 0}\alpha_1(\mathbf{p})=
\left\{ \begin{array}{cc}a_1(\mathbf{p}) & \omega_p<\omega_c \\ a_2(\mathbf{p}) &
\omega_p>\omega_c \end{array}\right.
\qquad\mbox{and}\qquad
\lim_{\mu\to 0}\alpha_2(\mathbf{p})=  
\left\{ \begin{array}{cc}a_1(\mathbf{p}) & \omega_p>\omega_c \\ a_2(\mathbf{p}) &
\omega_p<\omega_c \end{array}\right.
\end{equation} 
Eq. (\ref{high density limit}) tells us that when neutrinos are released from the
neutrinosphere in a supernova, the system occupies one of the eigenstates of the
Hamiltonian in Eq. (\ref{lagrange}) because the neutrino density is initially very high
and all neutrinos emerge in flavor eigenstates. Under the adiabatic evolution conditions,
the system stays in the same eigenstate but $\alpha_1$ and $\alpha_2$ slowly evolve from
flavor to the mass basis. According to Eq. (\ref{vacuum limit}), all neutrinos are
converted from flavor basis into mass basis by the time the neutrino density drops to
zero. 

\begin{figure}
\centering
\includegraphics[width=0.55\textwidth]{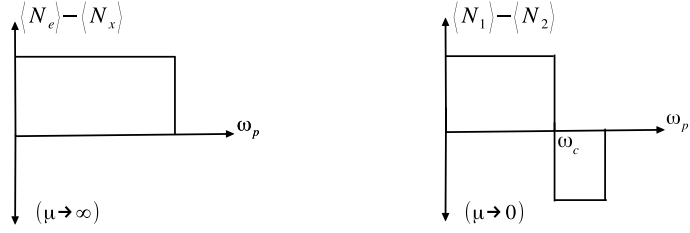}
\caption{Adiabatic evolution of the state given in Eq. (\ref{State}) from a region of
high neutrino density ($V\to 0$, $\mu\to\infty$) to a region of low neutrino density
($V\to \infty$, $\mu\to 0$).}
\label{evolution}
\end{figure}

As an example, let us consider an initial state which simply consists of electron
neutrinos with a box spectrum as shown on the left hand side of  Fig. (\ref{evolution}).
According to Eq. (\ref{high density limit}), this distribution corresponds to the
eigenstate given in Eq. (\ref{State}) in the limit where $\mu\to \infty$. In the
opposite limit where $\mu\to 0$,  Eq. (\ref{vacuum limit}) tells us that the same eigenstate 
corresponds to the distribution seen on the right hand side of Fig. (\ref{evolution}).
Therefore those neutrinos which oscillate faster than a critical frequency evolve into the
second mass eigenstate whereas others evolve into the first mass eigenstate. 

At this point, we would like to leave the mean field approximation scheme and return to
the exact many-body Hamiltonian given in Eq. (\ref{total hamiltonian}) and its dynamical
symmetries represented by the constants of motion given in Eq. (\ref{Invariants}).
Although we have worked with two mixing flavors so far, it is easy to show that 
similar symmetries exist for three mixing flavors as well. In order to describe three
neutrino flavors, one should first generalize the concept of neutrino isospin. 
For this purpose, we introduce the operators
\begin{equation}
\label{3isospin}
\mathcal{T}_{ij}(\mathbf{p})=a_i^\dagger(\mathbf{p})a_j(\mathbf{p})
\end{equation}
for $i,j=1,2,3$. These operators generalize the mass isospin defined in
Eq. (\ref{Mass Isospin Operators}). 
We use curly letters to denote the components of the operators in mass basis to comply with
our earlier convention. Note that we also generalize the summation convention introduced
in Eq. (\ref{Totals}) to the three flavor case. For example, $\mathcal{T}_{ij}$
represents the $ij$ component of the generalization of the total isospin of all neutrinos.
These operators now obey the $SU(3)$ commutation relations:
\begin{equation}
[\mathcal{T}_{ij}(\mathbf{p}),\mathcal{T}_{kl}(\mathbf{q})]=
\delta^3(\mathbf{p}-\mathbf{q})\left(\delta_{kj}\mathcal{T}_{il}(\mathbf{p}) 
-\delta_{il}\mathcal{T}_{kj}(\mathbf{p})\right)
\end{equation}
The Hamiltonian which describes the self interactions and the vacuum oscillations of
neutrinos is given by
\begin{equation}
\label{3 flavor hamiltonian}
H=-\sum_p 
\left(\frac{\delta m^2_{21}+\delta m^2_{31}}{6p}\mathcal{T}_{11}(p)+
\frac{\delta m^2_{12}+\delta m^2_{32}}{6p}\mathcal{T}_{22}(p) 
+\frac{\delta m^2_{13}+\delta m^2_{23}}{6p}\mathcal{T}_{33}(p)\right) 
+\frac{\mu}{2}\sum_{i,j=1}^3\mathcal{T}_{ij}\mathcal{T}_{ji}
\end{equation}
where $\delta m^2_{ij}=m_i^2-m_j^2$ and the single angle approximation is again adopted in
describing the neutrino-neutrino interactions. This Hamiltonian reduces to the one given
in Eq. (\ref{total hamiltonian}) up to a trace term if it is restricted to a two flavor
subspace. It is a simple exercise in algebra to show that the operators  
\begin{equation}
\label{3 flavor invariants}
h_p=-\left(
\frac{\delta m^2_{21}+\delta m^2_{31}}{6p}\mathcal{T}_{11}(p)+
\frac{\delta m^2_{12}+\delta m^2_{32}}{6p}\mathcal{T}_{22}(p)+
\frac{\delta m^2_{13}+\delta m^2_{23}}{6p}\mathcal{T}_{33}(p) 
\right)
+\mu\sum_{q(\neq p)}\sum_{i,j=1}^3
\frac{\mathcal{T}_{ij}(p)\mathcal{T}_{ji}(q)}{\frac{1}{2p}-\frac{1}{2q}}
\end{equation}
are constants of motion of the Hamiltonian given in Eq. (\ref{3 flavor hamiltonian}),
i.e., they obey 
\begin{equation} 
\left[{h}_p,{h}_q\right]=0\qquad\mbox{and}\qquad\left[{H},{h}_p\right]=0 
\qquad\mbox{for all $p$.}\qquad
\end{equation} 
If restricted to a two flavor subspace, these constants of motion are equivalent to the
ones given in Eq. (\ref{Invariants}) up to a trace term and an overall multiplicative
constant. 

We emphasize that a study of self interacting neutrinos as a many-body
system with its underlying symmetries can lead us to a simple understanding of its
collective modes of behavior. To this end, we presented the invariants representing the
dynamical symmetries in the exact many-body formalism. In the two flavor mixing scheme, we
used an example in which a single spectral split can be simply viewed as an avoided level
crossing of the mean field Hamiltonian if the value of a many-body constant of motion is
fixed with a Lagrange multiplier. We showed that the many-body Hamiltonian describing
three mixing flavors with self interactions also has the same dynamical symmetries. Whether
these symmetries can be used to explain multiple spectral splits of neutrinos for two or
three mixing flavors and whether they lead to other, possibly more interesting collective
behavior modes are the subjects of our ongoing research and will be discussed elsewhere. 

\vspace*{3mm}
\noindent We thank to B. Szczerbinska for organizing CETUP* 2013. We also thank to the
participants for their valuable comments.  
This work was supported 
in part by the Scientific and Technological Research Council of Turkey
(T{\"{U}}B{\.{I}}TAK) under project number 112T952,
in part by the U.S. National Science Foundation Grant No.
PHY-1205024, in part by the University of Wisconsin Research Committee with funds
granted by the Wisconsin Alumni Research Foundation, 
and in part by Grants-in-Aid for Scientific Research of JSPS (20105004, 24340060) of the
Ministry of Education, Culture, Sports, Science and Technology of Japan.

\bibliographystyle{aipproc} 
\bibliography{mn-jour,biblio}
\end{document}